# Can Commercial Testing Automation Tools Work for IoT? A Case Study of Selenium and Node-Red

Neenu Varghese and Roopak Sinha
*IT & Software Engineering*
*Auckland University of Technology*
*Auckland, New Zealand*
byd8050@autuni.ac.nz, roopak.sinha@aut.ac.nz

**Abstract**

***Background:*** *Testing IoT software is challenging due to large scale, volume of data and heterogeneity. Testing automation is a much-needed feature in the domain.* ***Aims****: The first goal of this research is to explore the requirements and challenges of IoT testing automation. The second goal is to integrate testing automation tools used in commercial software into the IoT context.* ***Method:*** *A systematic literature review is carried out to elicit requirements for testing automation in IoT. A design science approach is followed to build a testing automation tool for IoT applications written in the Node-Red platform, using the commercial testing automation tool Selenium. The resulting framework uses the Selenium Web Driver for browser-based testing automation for IoT applications.* ***Results:*** *The proposed framework has been functionally tested on multiple browsers with preliminary evaluation on maintainability, browser capability and comprehensiveness.* ***Conclusions:*** *The use of commercial tools for testing automation in IoT is feasible. However, major challenges like high data volumes and parallel transmission and processing of data need to be addressed comprehensively for complete integration.*

## 1. INTRODUCTION

Internet of Things (IoT) is a vast network of interconnected devices and systems. IoT software applications process large amounts of data with support from the underlying networking and hardware infrastructure. An important challenge in IoT is maintaining the quality of services provided over a heterogeneous IoT network. Service quality can vary due to variation in the frequency and timing of data being transported over the network. IoT allows real-time decision-making over distributed data streams. For instance, a multi-media service may use an IoT application to receive failure alerts and take necessary actions to recover from failures. Compared to traditional systems, an IoT based solution can effect and track such activities in great detail with the help of data from sensors and monitoring devices. This data can be stored and mined to create refinements to the system.

The testing of IoT applications revolves around the networks, device, platform, security, processors, standards, and operating system. As IoT consists of various components such as sensors, application, network and data centre in which each component is capable of different functionalities which require different types of testing. Among these components, the application component is the only area which applies to all types of testing such as functionality, security, usability, operational stability, services, performance and compatibility testing. In the research by [1], the testing of IoT systems revolves around overall quality of communication protocol implementation and comparison of quality. They have tried to prove the hypothesis experimentally with open source protocol brokers.

Several studies have considered the issue of automated testing of IoT applications, which is extremely hard due to their highly dynamic and heterogeneous nature. Automation can improve the management and business processes for IoT systems. Similarly, automating the testing of IoT systems and application can play an equally important role. The main challenges of testing IoT system applications lie in triggering functionality, fast autonomous response, operational optimisation and most importantly, integration. In [2], a prediction engine for the utilisation of resource had been developed to trigger the adaptation of IoT systems. A smart test framework over an emulated IoT platform is used to simulate real data traffic within a heterogeneous system to test applications. Machine learning algorithms are subsequently used for predicting resource utilisation while maintaining the desired quality of service. Limitations of such works involve the effort required to simulate data precisely, such as movement data like in wearable devices. In another work [3], a service-based testing framework called IoT-TaaS is proposed to explore issues related to costs, scalability and coordination of traditional ways of testing, implementing and designing

IoT systems. This method provides interoperability testing of remote systems, validation testing for semantics, conformance testing for scalability. However, a survey of related literature, provided in Sec. 3, reveals that testing automation is not commonplace in IoT.

This paper provides an initial step towards testing automation in IoT at the same level as enterprise software. We study Selenium [4], the de-facto testing automation tool in enterprise and commercial software, for its suitability in the IoT context. Specifically, we explore the following research questions:

**RQ1** What are the challenges of automated testing of IoT software applications?

**RQ2** What are the requirements for testing software complying with the IoT reference architecture?

**RQ3** To what extent does the contemporary test automation tool Selenium meet the requirements identified in RQ2?

The rest of this paper is organised as follows. Sec. 2 defines important background concepts used throughout the rest of this paper. RQ1 and RQ2 are answered primarily through a systematic literature review, reported in Sec. 3. RQ3 is answered following a design science research methodology [5] in which an integration of Selenium with the IoT simulator environment Node-RED [6] is carried out to test Selenium's feasibility in the IoT context. This involves finding dependencies between the Selenium Web Driver and Node-RED and implementing a unit-testing framework for this setting. The framework design includes scheduling test execution, build management, integration with a reporting mechanism and setting up continuous integration and continuous delivery. The implemented framework is evaluated based on metrics identified during the systematic literature review and coverage of IoT testing requirements. Other tests included regression testing for ensuring stability, and an analysis of usability, performance and test execution speed.

## 2. BACKGROUND

Internet of Things (IoT) is a set of patterns which contains smart devices collaborating with virtual and physical resources within a connected network. The reference architecture model of IoT [7] allows us to understand the structure of IoT systems in order to elicit requirements on testing (to answer RQ2).

The main requirements of applications complying with IoT reference architecture are:

- *Device management* including tasks like the ability to disconnect stolen devices, updating the system credentials, locating lost devices, updating software in each device, erasing secure data from stolen devices, enable and disable hardware functionalities, etc. [7].

- *Collection of data and analysis* including handling large amount of data from these sensors and actuators, and to process, store and analyse this data using highly scalable storage capacity.

- *Security*, including privacy, access management, confidentiality and availability [7].

- *Scalability* to handle varying and large volumes of data simultaneously through features like elastic deployment.

- *Connectivity and Communication* challenges like choosing networking protocols (other than HTTP) that provide a balance between performance, security, and power consumption [7].

The IoT Reference Architecture takes a layered approach to manage high complexity. The different layers are device layer, communication layer, bus or aggregation layer, event processing and analytics layer and the top layer as application layer which handles external and client communications [7]. The application layer is the highest layer of the architecture. The transmission, processing and analysis of data is performed on the application layer [8]. In this research, the testing framework is implemented on the IoT application system and as a result, we are focusing more on the application layer of the IoT architecture.

*Testing* involves ensuring that an IoT application meets its desired purpose and functionality. Test cases can be executed objectively to improve the system, understand risks and refine the related business processes [1]. White-box testing or clear box testing exploits the internal structure of the implementation design of the software. Black-box testing treats the software as a black box and focuses on stimulating it with carefully chosen input combinations and observing the resulting outputs. According to [1], both white and black-box testing can be used during system development to ensure the right balance of coupling between test cases and the software being tested. Black-box testing can be reused across the implementation stages and hence increase the interoperability of the system. Testing is essential in ensuring not just correct functionality, but also the quality of an IoT application. Such qualities include timeliness, reliability and security.

Testing automation automates the process of testing, allowing this labour-intensive task to become more scalable and feasible for larger systems. Selenium [12] is arguably the most commonly-used testing automation tool used in commercial and enterprise software nowadays. As a web-based tool, Selenium provides several benefits like being open-source and gree to use, availably integration with DevOps, support for multiple browsers, operating systems, programming languages, hosting simulated and real mobile devices, and the ability to run multiple test cases in parallel [13].

## 3. RELATED WORKS

We conducted a systematic literature review (SLR) [14] to answer RQ1 and RQ2. The search terms were derived by breaking the research questions into terms and then connecting them logically using AND and OR operators. The final search term used for this SLR was `"IoT" OR "Internet of Things" AND "Automated Testing" AND "Challenges" AND "Requirements" OR "Automated Testing` using Selenium" OR "IoT Reference Architecture" OR "IoT Testing framework" OR "Testing techniques in IoT" OR "IoT testing using open source tool" OR "IoT testing methodology". We searched for primary studies published from 2010 onwards and published in peer-reviewed venues. The databases used for the search were ACM Digital Library, Science Direct and IEEEXplore.

Screening of papers was done by creating inclusion and exclusion criteria. Papers that focused on topics like IoT software, testing related challenges or technology in IoT, requirements of IoT systems and Selenium based automation were included. All duplicated results, or studies on hardware or networking aspects of IoT, or those deemed to not provide significant technical insights into testing of IoT software were excluded. This screening was conducted by reading paper abstracts along with a quick scan of the whole document. A total of 17 papers was finally selected for this review. Each of the 17 papers were read fully to extract information relevant to answering the research questions. The main findings of the SLR are described below.

Several works discussed testing related challenges in IoT software. In [15], some challenges of IoT testing are mentioned, including that testing in real-time systems is harder than testing interaction with web browsers. This paper does not mention testing automation. This work mentions different types of testing considerations are mentioned which only partially answers RQ2. The main challenges of IoT regardless of the type of testing are discussed in [16]. Other challenges include availability, reliability, mobility, performance, scalability, interoperability and management [17].

Several studies mention IoT architecture and system specifications. As per [18] an IoT reference architecture is essential since it defines the design and building blocks for applications. IoT ARM and architecture by WSO2 are two recent reference architectures of IoT [18]. In IoT, the ARM architecture provides different perspectives and architectural views for the construction of the IoT platform, following a top-down method of baselining the models. In WSO2, a bottom-up approach utilising both server-side and cloud architectures are used to create interaction between devices. Both architectures focus on the requirements of interoperability, scalability, privacy, security, integrity and performance. Two existing IoT architectures: SoA-based and three layer architecture (Application, Network and Perception layer), are briefly discussed in [8].

Commercially used tools like Selenium cannot be used directly for IoT testing. IoT software is different and more complex when compared to traditional software. Challenges include time-accurate simulation of sensors and other devices, the large number of nodes, and inherent heterogeneity. According to [15], the types of testing required for IoT applications beyond functional testing include testing for *usability*, *scalability*, *reliability*, *compatibility*, *data integration*, and *performance* testing.

We find three key metrics to compare the different testing frameworks we surveyed. A comparison of existing testing frameworks for IoT software is shown in Tab. I. The first metric, *test environment*, refers to the extent to which the environment in which a test is performed closely relates to a real environment [15]. Large scale and heterogeneity can make IoT testing difficult. As an example, in [2] a smart testing framework is implemented on EMU-IoT. This is a containerised environment which is modeled with a heterogeneous network so that combination of large and small data are produced as needed. The IoTTest framework [9] is a specialized testing based on JUnit framework for testing E2E device and simulators. The application of combining the process of unit simulation and integration testing helps the IoT testing to reduce the cost and streamline the testing process. The testing performed in [10] uses a model communication network for information exchange between components. In [11], pattern based IoT testing called Izinto for the integration testing of IoT systems is proposed.

The second important metric of comparison is the *purpose* of the testing. The main issue for testing an IoT system is its heterogeneous network and the fluctuation of large volumes of data occurring at the same time. The aim of the work presented in [2] is maintaining the quality of service in heterogeneous networks and increase the accuracy of the whole system with the help of predictive analysis. The unit integration or E2E integration when performed alone in an heterogeneous network has certain limitations and challenges. In [9], test orchestration, device library and device simulation are the primary objectives. This makes the framework capable of terminating the test chain if it fails. It also handles connection issues and orchestrates unit integration modules with E2E chain. The approach in [3] supports distributed and automated testing of IoT devices with test systems in remote locations for resolving the issues with scalability, cost and coordination of traditional software testing methods. A full-scale experiment on a fragment of an IoT communication network was done in [10] in order to select the best application protocol for an ideal IoT system. The validation of generated test patterns which are implemented by the framework of [11] was done on an application scenario in AAL domain.

TABLE I. SYSTEMATIC LITERATURE REVIEW RESULTS

| Work | Test Environment | Purpose | Test Methodology |
|---|---|---|---|
| Smart Testing Framework [2] | Heterogeneous network of Emulated IoT (EMU-IoT) | Maintaining quality of service (QoS) Identifying bottlenecks | Resource utilization prediction engine |
| IoTTest Framework [9] | End-to-End IoT and Simulators | Test orchestration Device simulation Device library | Combined principle of simulation with unit integration and E2E integration testing |
| IoT Application Protocol Testing [10] | Model communication network | Select the application protocol for IoT, comparing the average response time (RTT) in MQTT, CoAP, and HTTP protocols | Comparing server response time to the request |
| IoT Testing as a Service (IoT-TaaS) [3] | IoT binding protocol and device | Resolve constraints on coordination, costs, and scalability issues software testing. in the context of standards-based development of IoT devices | Remote distributed interoperability testing, scalable automated conformance testing and semantics validation testing |
| Izinto, a pattern based IoT testing framework [11] | Concrete application scenario in the domain of Ambient Assisted Living (AAL) | Test IoT solutions in an automated manner and with reduced effort, by implementing a set of test patterns out-of-the-box | Identify a set of test strategies to test recurring behaviours of IoT systems which can be described as test patterns |

The third metric of comparison is *test methodology*. In [2], a resource utilisation engine is used to identify bottlenecks in IoT networks by studying usage patterns. A hybrid approach combining the idea of simulation, unit integration and E2E integration is used in [9]. In [3], a standard-based testing approach is taken to reduce the constraints of IoT testing considering interoperability testing, validation testing and conformance testing for an integrated system. A comparison mechanism was used in [10] to study differences in server response times for different protocols. In the Izinto framework, a set of test strategies are identified to validate the recurring behaviours in an IoT system which are captured as test patterns [11].

## 4. DESIGN AND IMPLEMENTATION

**A. Case Study: Smart Metering**
A smart metering case study was used to build the testing framework. Smart meters simplify billing processes, provide continuous tracking and ensure up-to-date and accurate measurements [19]. There are several existing simulators available to simulate different layers of architecture. For instance, Qualnet and OMNET++ are networks simulating tools mostly used to examine network interference and device placement planning [20]. In our work, we have used Node-Red, a programming tool for IoT development which wires hardware devices, online services and APIs. It is a browser-based editor platform makes it easy to deploy the developed flows into run time on a single click. Function nodes are created using JavaScript in respective text editors.

The smart metering system is simulated using Node-Red in order to recreate different scenarios and use cases by avoiding the involvement of expensive real sensors. Moreover, the simulator can be used with a combination of real sensors. The two main goals of the system are to test the application which requires sensor data and to evaluate deployment feasibility prior to the deployment in real equipment.

The flows in Node-red are managed by different types of nodes as shown in Fig. 1. Each node has a well-defined function or purpose. Each node receives data, processes it and potentially produces additional data for consumption by downstream nodes. The network is responsible for the flow of data between nodes. The simulated smart metering system is designed to get the alert notification mail to the end-user once the input data value (fuel unit) has crossed the limit. In this system, 3 separate mails are configured for different fuel levels *low level* (0 - 10 units), *warning level* (10-50 units) and *critical level* (50-100 units). Three function nodes are defined for each of the value levels. The fuel unit node acts as a substitute for a real sensor capturing

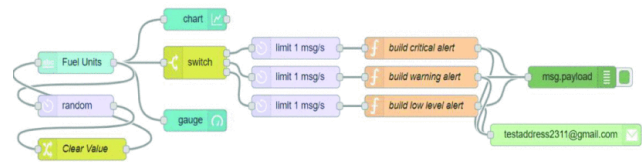
Fig. 1. Smart Metering System(AUT)

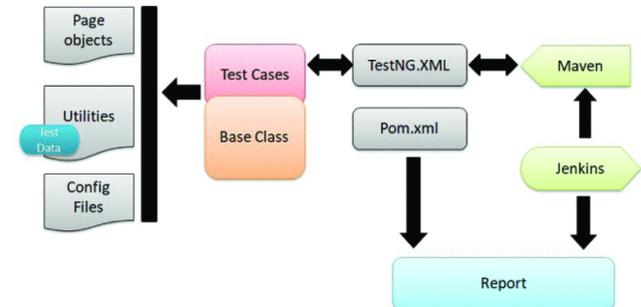
Fig. 2. Framework Architecture

the fuel-level reading from the fuel tank or fuel meter. Once data is received, it is passed to the switch node in order to direct the output to different parts of the flow depending on the input. Switch node has 3 output terminals directing to low level, warning level and critical level function nodes. The delay node limits the messages passing through the modes in terms of seconds, minutes, hours or day in order to receive the alert accordingly. The function nodes build the alert mails which are JavaScript function blocks to run against the messages being received by the node. The topic and body of the mail are defined and returned as a message object. The email node sends the message from the function in the form of mail to the email address which is set up. The `msg.payload` node acts as a debug node to get the run-time log. The random and clear value nodes together clear the input data after a random delay in order to accept the new data coming.

**B. Selenium-based Testing Automation Framework**
We designed an automated testing framework based on Selenium for Node-Red applications. The Eclipse IDE was used for framework development in Java. Fig. 2 shows the overall organization of the framework. Integration with Maven and Jenkins allows Selenium to interface with third-party tools for continuous integration and delivery.

Fig. 3 shows how different components of the framework interact. The test project consists of different test packages which are represented as components. The classes of each package are represented as attributes in each component. Here, the test project contains packages like page objects, utilities, test data and test cases. Moreover, the test project contains log4j listener file, driver file with exe file of different browsers for cross-browser testing, `testNG.xml` for reporting and integration with TestNG, `pom.xml` contains configuration information, project information for Maven such as construction directory, source directory, dependency, test source directory, goals and plugins. Maven is the build management tool based on the concept of page object model (POM). Once the test cases are created, they are executed using `testNG.xml` and `pom.xml`. Maven's batch feature is used to run the test suite using the command line instead of running through IDE. The test suite is run through Jenkins for continuous integration.

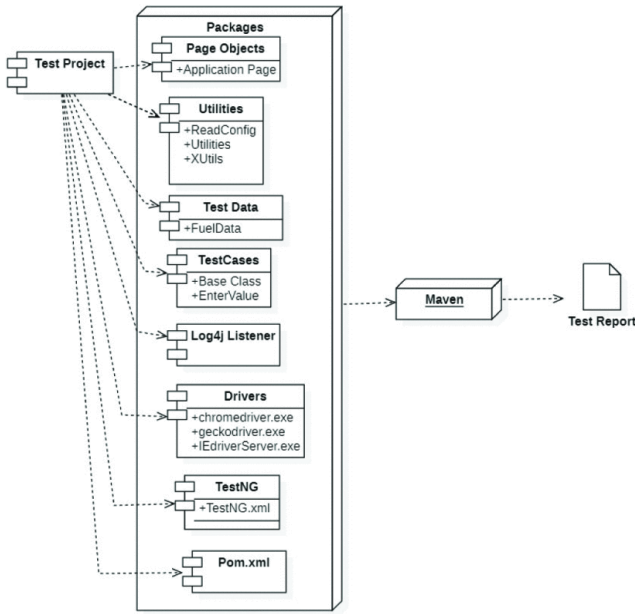

Fig. 3. Component Diagram

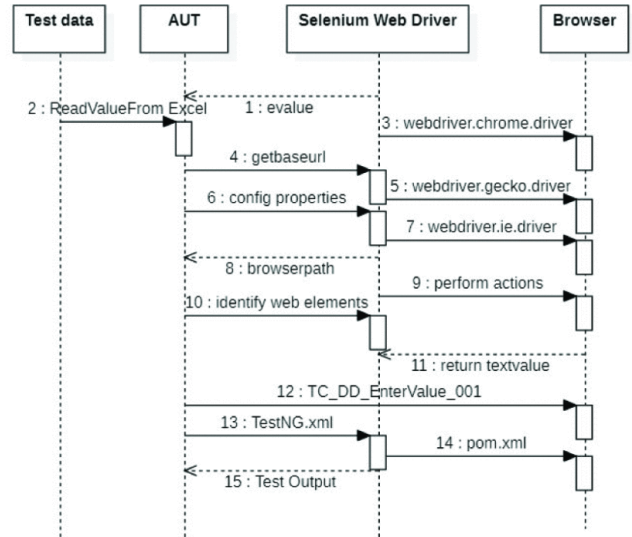

Fig. 4. Sequence Diagram

Fig. 4 shows the interactions between the components of the automated testing framework. The test data is read from an excel spreadsheet which is managed by XLUtils class. Application Under Test (AUT) is available to all web browsers with dashboard UI view of Node-Red application. Selenium Web Driver interfaces and creates a connection between the application under test and the web browser. Driver exe files of different browsers are stored in the driver folder which is configured using `ReadConfig` class. The web elements identified as `xpaths` are defined in the homepage class and called using web driver functions along with the actions to perform on each element.

## 5. EXPERIMENTAL RESULTS

The proposed framework was tested over a smart metering system, and tests were focused on the exchange of data between different devices, and nodes in real-time.

**Functional Testing:** This is black or white box testing applied to any Application Under Test (AUT). The ultimate aim is to ensure that the end-to-end functionality of the system conforms to requirements without throwing any errors. In our framework, this is done by loading the test cases onto a web page enabled by the Selenium Web Driver. Expected outputs to input vectors are then used to automatically check if tests are successful. Tests are executed, and results are reported using the framework, following the process shown in Fig. 4. Errors are tracked for each failed test case and recorded with screenshots. Fig. 5 shows the test HTML report generated in the Selenium framework using the "Extend Reports" feature.

**Cross-browser testing:** The framework was tested with Chrome, Internet Explorer, Safari and Firefox browsers. For executing the test suite in multiple browsers, Selenium Webdriver is integrated with the "TestNG" framework. The file TestNG.xml is defined with separate test tags and parameters for each browser. Upon passing the parameter value from XML file to test case, Web Drivers are created for each browser automatically. Fig. 6 shows the test result trend for cross-browser testing obtained on running test cases through Jenkins.

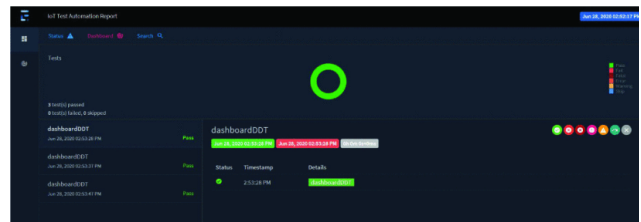

Fig. 5. Functional Testing Automation Report

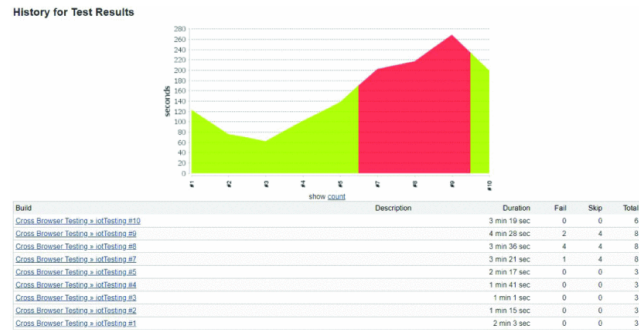

Fig. 6. Cross-browser test result trend in Jenkins

**Load Testing:** During peak or heavy load conditions, multiple tests are run simultaneously to understand system performance and functional suitability. Selenium is not optimised for load/stress testing, especially if the browsers are running locally. While running more than 1000 browser tabs, each hosting an individual test, at the same time causes the system to crash. The main bottleneck is limited RAM, CPU capacity and network bandwidth, which make it extremely hard to ensure real-time interaction between system events and tests. We used JMeter, executing independently, for load and performance testing of the system. In JMeter, a group of virtual users are simulated to send requests to the target server (the automation testing framework) and captures performance statistics. Fig. 7 and Fig. 8 shows the result table and summary of load testing in the simulated IoT application using JMeter. The test run is loaded with 200 threads(virtual users) and three loop count(number of iterations for each thread).

Fig. 7. Load Testing result table in Jmeter

Fig. 8. Load Testing result summary in Jmeter

The varied tests executed on the case study show the usability of Selenium in testing automation of Node-Red applications. The prototype implementation works well for unit testing and can be scaled for integration tests. However, high volume and distribution mean that the interfaces between the testing automation system and the AUT become the key bottlenecks. There is scope to address these issues by optimising the flow of events between these components.

## 6. CONCLUSIONS AND FUTURE WORK

We design and develop a Selenium based testing automation tool for IoT systems developed using Node-Red. Tested over a case study of smart metering, we find that the proposed framework confirms the feasibility of using Selenium and other commercial testing automation tools in the IoT context. At the same time, a systematic literature review and the experimentation on the case study highlights some additional challenges that need to be addressed in the future, such as the real-time nature and high volumes of data being generated in a complex IoT system.